\documentclass[article,preprint,showpacs,amsfonts]{revtex4}
\usepackage{indentfirst}
\usepackage{amsmath,amssymb}
\def\ra{\rangle}
\def\la{\langle}

\begin{document}

\title{Trace distance measure of coherence for a class of qudit states}
\author{Zhen Wang}
\affiliation {Department of Mathematics, Jining University,
Qufu 273155, China}
\author{Yan-Ling Wang}
\affiliation {Department of Mathematics, South China University of Technology,
Guangzhou 510640, China}
\author{Zhi-Xi Wang}
\affiliation {School of Mathematical Sciences, Capital Normal
University, Beijing 100048, China}

\begin{abstract}
Recently, trace distance measure of coherence has been proposed for characterizing the coherence of a given quantum
state. However, it seems difficult to estimate the optimal incoherent state for high dimensional states.
An explicit expression for the trace distance measure of coherence for a class of qudit states is provided. We show that
the closest incoherent state to the class of qudit states is just the diagonal matrix obtained from it by deleting
all off-diagonal elements. It is also shown that the measure of coherence induced by trace distance can act as a measure of coherence
for this class of qudit states with some restrictions on the strictly incoherent operations.

\noindent{\it Keywords}: trace distance of coherence, closest incoherent state, strictly incoherent operation
\end{abstract}

\pacs{03.67.Mn, 03.67.Ac, 03.65.Ta}

\maketitle
It is known that coherence or superposition is one of the most fundamental quantum features. As a key feature of quantum mechanics, quantum coherence plays more and more important role in physics. The importance of quantum coherence is introduced by wave particle duality in physical phenomena such as nanoscale physics \cite{karlstrom}, quantum thermodynamics \cite{aberg2,lostaglio1,lostaglio2} and low temperature thermodynamics \cite{nara}. Numerous works has been undertaken to formulate quantum coherence theoretically \cite{glauber,sudarshan,luo,aberg,monras,baumgratz,aberg2,girolami}. Only very recently, the framework of the quantification of quantum coherence using methods of quantum information science has been proposed and subsequently developed \cite{baumgratz,girolami,pires,streltsov}. Baumgratz {\it et. al.} constructed a rigorous framework of quantifying coherence and introduced several measures of coherence based on well-behaved metrics including relative entropy, $l_p$ norm, trace norm and fidelity \cite{baumgratz}. They also proposed the following conditions which a proper measure of the coherence $C$ must satisfy:

(C1)\quad $C(\rho)\geq 0$ for all quantum states $\rho$ and it should be zero for all incoherent states.

(C2a)\quad Monotonicity under all incoherent channels $\Lambda$, i.e.
$C(\rho)\geq C(\Lambda(\rho))$, where $\Lambda$ acts as $\Lambda(\rho)=\sum_n K_n\rho K_n^\dag,\ \{K_n\}$ are a series of Kraus operators which satisfy
$\sum_n K_n^\dag K_n=\mathbb{I}$ and $K_n {\cal I}K_n^\dag\subset{\cal I}$.

(C2b)\quad Monotonicity under selective incoherent channels: $C(\rho)\geq \sum_n p_n C(\rho_n)$, where
$\rho_n=(K_n\rho K_n^\dag)/p_n,\ p_n={\rm tr}(K_n\rho K_n^\dag)$ for all $\{K_n\}$ with $\sum_n K_n^\dag K_n=\mathbb{I}$ and
$K_n {\cal I}K_n^\dag\subset{\cal I}$.

(C3)\quad Nonincreasing under convex mixing of quantum states, i.e. $\sum_n p_n C(\rho_n)\geq C(\sum_n p_n\rho_n)$ for any ensemble $\{p_n,\ \rho_n\}$.

According to the above statement, one must find its closest incoherent state if one want to quantify the coherence of a quantum state. As shown in \cite{baumgratz}, the closest incoherent state for any state $\rho$ is the state which is obtained from $\rho$ by deleting all off-diagonal elements for the relative entropy and $l_1$ norm measure of coherence. In addition, it is also shown that both measures of coherence satisfy the conditions (C1)-(C3) which ensure that they are the proper measures of coherence. However, it has been illustrated that neither the fidelity nor the Hilbert-Schmidt distance is coherence measure since they violate the condition (C2b) \cite{shao,rana}. But nonetheless, it has not been testified whether the measure of coherence induced by trace distance can act as a good measure of coherence because it is difficult to estimate the optimal incoherent state for general high dimensional states even qutrit states. In this paper, we will present some progress finding the closest incoherent state for special qudit states to quantify the trace distance measure of coherence. Since the trace distance measure of coherence fulfils the conditions (C1)-(C3) \cite{baumgratz}, we will further investigate whether the trace distance measure of coherence obeys the condition (C2b) for these special qudit states.

Recall that the trace distance measure of coherence is formally defined as
$$
C_{\rm tr}(\rho):=\min_{\delta\in \cal{I}}\|\rho-\delta\|_1=\min_{\delta\in \cal{I}}{\rm tr}|\rho-\delta|,
$$
where ${\cal I}$ is the set of incoherent states, that are diagonal in a fixed basis. In order to obtain the trace norm of $\rho-\delta$, one must solve the eigenvalues of $\rho-\delta$. However, the complex expressions of the eigenvalues of the density matrices for quantum system show that in general the closest state $\delta$ cannot be determined explicitly, even if the dimension of $\rho$ is small. In \cite{shao}, Shao {\it et. al.} showed that $C_{\rm tr}(\rho)$ has the same form of expression
as the $l_1$ norm of coherence for the one-qubit case. The result is extended to general matrices instead of states is to get rid of the positivity
and the unit trace conditions\cite{rana}. In addition, the closest diagonal matrix to the X states and three classes of special qutrit states $\rho_X$ is given by the diagonal matrix to $X$ and $\rho_X$, respectively\cite{shao,rana}. Recently, the closest incoherent state to a given pure state is characterized under the trace distance measure of coherence by Chen {et. al.} in \cite{chen}. However, the optimization for qudit states is still very messy as it seems to be more difficult to express eigenvalues of general qudit states.

Inspired by \cite{rana}, we find the analytic form of trace distance measure of coherence for a class of qudit states
\begin{eqnarray}\label{rho}
\rho=\left(
\begin{array}{cccc}
x_1&a&\cdots&a\\
a&x_2&\cdots&a\\
&\cdots&\cdots&\\
a&a&\cdots&x_d\\
\end{array} \right),
\end{eqnarray}
where $\displaystyle\sum_{i=1}^dx_i=1$ and all principle minors of $A$ are nonnegative which ensure that $\rho$
is a quantum state. In the below, we will analyze the trace distance measure of coherence for this class of qudit states.
Before that, we need recall the below lemma.

{\bf Lemma 1}\cite{horn2}\quad Let $A=(a_{ij})\in M_{st}$ be given and $q=\min\{m,n\}$. Let $a\equiv(a_{11},a_{22},\cdots,a_{qq})^T\in C^q$ and
$\sigma(A)\equiv(\sigma_1,\sigma_2,\cdots,\sigma_q)^T\in R^q$ denote the vectors of main diagonal entries and singular values of $A$, respectively.
Let $|a_{[1]}|\geq |a_{[2]}|\geq\cdots\geq |a_{[q]}|$ and $\sigma_{[1]}\geq\sigma_{[2]}\geq\cdots\geq\sigma_{[q]}$ denote the absolutely decreasingly ordered main diagonal entries and decreasingly ordered singular values of $A$, respectively, then
\begin{equation}\label{lemma}
|a_{[1]}|+|a_{[2]}|+\cdots+|a_{[i]}|\leq\sigma_{[1]}+\sigma_{[2]}+\cdots+\sigma_{[i]}
\end{equation}
for $i=1,\ 2,\ \cdots,\ q$.

Specially, Lemma 1 implies that the maximal absolutely main diagonal entry is less than the maximal singular value for any matrix $A$, i.e., $|a_{[1]}|\leq\sigma_{[1]}$.

Although it is intuitively expected that the closest incoherent state for a quantum state $\rho$ is the diagonal matrix ${\rm diag}(\rho)$, in general it is not true for high dimensional states even pure states. Fortunately, we find that it is true for the qudit states $(\ref{rho})$.

{\bf Theorem 2}\quad Let $\rho$ be a $d\times d$ matrix in (\ref{rho}) and $\delta$ be its closest diagonal matrix in trace norm.
Then $C_{\rm tr}(\rho)=2(d-1)|a|$ and $\delta={\rm diag}(\rho)$.

{\bf Proof.}\quad Let $\delta=\displaystyle\sum_{i=1}^d\delta_i|i\ra\la i|$, where $\delta_1+\delta_2+\cdots+\delta_d=1$. It is not difficult to
find that $\rho-\delta$ is traceless. By use of the Sherman-Morrison-Woodbury formula for determinants \cite{horn}:
$$
\det(A+uv^T)=(1+v^TA^{-1}u)\det A,
$$
we have
\begin{equation}\label{det2}
\det[\lambda I-(\rho-\delta)]=(1-\displaystyle\sum_{i=1}^d\frac{a}{\lambda-x_i+\delta_i+a})\prod_{i=1}^d
(\lambda-x_i+\delta_i+a).
\end{equation}
Let $f(\lambda)=\det[\lambda I-(\rho-\delta)]$, obviously, $f(\lambda)$ can be considered as a polynomial function on $\lambda$, so $f(\lambda)$ is a continuous function at the interval $(-\infty,+\infty)$.

In order to use Taylor series expansion for real function $\frac{1}{1-x}$ about $x=0$, at first, assume $|\frac{x_i-\delta_i}{da}|<1$, i.e., $|x_i-\delta_i|<|da|$ for all $i=1,2,\cdots,d$. For convenience, let $y_i=x_i-\delta_i$ for $i=1,\ 2,\ \cdots, d$. Using the property that $\rho-\delta$ is traceless, we have
\begin{eqnarray*}
f((d-1)a)&=&\left(1-\sum_{i=1}^d\frac{a}{da-y_i}\right)\prod_{i=1}^d(da-y_i)\\
&=&\left(1-\frac{1}{d}\sum_{i=1}^d\frac{1}{1-\frac{y_i}{da}}\right)\prod_{i=1}^d(da-y_i)\\
&=&\left(1-\frac{1}{d}\sum_{i=1}^d(1+\frac{y_i}{da}+(\frac{y_i}{da})^2+\cdots)\right)\prod_{i=1}^d(da-y_i)\\
&=&\left(-\frac{1}{d}\sum_{i=1}^d\left((\frac{y_i}{da})^2+(\frac{y_i}{da})^3 +\cdots\right)\right)\prod_{i=1}^d(da-y_i),
\end{eqnarray*}
where we use Taylor series expansion of $\frac{1}{1-\frac{y_i}{da}}$ for $|\frac{y_i}{da}|<1$ in the third equality.
Let
$$
g(y_1,y_2,\cdots,y_d)=-\frac{1}{d}\sum_{i=1}^d\left((\frac{y_i}{da})^2+(\frac{y_i}{da})^3 +\cdots\right),
$$
then
$$
f((d-1)a)=g(y_1,y_2,\cdots,y_d)\prod_{i=1}^d(da-y_i).
$$
If $y_i=0$ or $x_i=\delta_i,$ for all $i=1,\ 2,\ \cdots,\ d$, we have $g(y_1,y_2,\cdots,y_d)=0$, thus $f((d-1)a)=0$. It implies that $(d-1)a$ is an eigenvalue of $\rho-\delta$ when $x_i=\delta_i$ for all $i$. Obviously, ${\rm tr}|\rho-\delta|=2(d-1)|a|$ when $x_i=\delta_i$ for all $i=1,\ 2,\ \cdots,\ d$. In fact, $\delta={\rm diag}(\rho)$ if $x_i=\delta_i$ for all $i=1,\ 2,\ \cdots,\ d$. That is, ${\rm tr}|\rho-\delta|=2(d-1)|a|$ when $\delta={\rm diag}(\rho)$. In addition, if there exists $y_i\neq0$ for some $i$, we have $g(y_1,y_2,\cdots,y_d)<0$. Now we solve the eigenvalue of $\rho-\delta$ when $|y_i|<|da|$ for all $i=1,\ 2,\ \cdots,\ d$ and $y_i\neq0$ for some $i$.

If $a>0$, then $\prod_{i=1}^d(da-y_i)>0$, therefore $f((d-1)a)<0$. Meanwhile, $f(+\infty)>0$. Applying Bolzano's theorem \cite{apostol} to $f$, we have $f(z)=0$ for some $z\in((d-1)a,+\infty)$. It shows that $z$ is an eigenvalue of $\rho-\delta$. This means that there exists an eigenvalue $z>(d-1)a$ for $\rho-\delta$.

Suppose $\lambda_1\geq\lambda_2\geq\cdots\geq\lambda_k\geq0>\lambda_{k+1}\geq\cdots\geq\lambda_d$ be the decreasingly ordered eigenvalues of $\rho-\delta$.
Thus we get
$$
{\rm tr}|\rho-\delta|=2(\lambda_1+\cdots+\lambda_k)\geq2z>2(d-1)a=2(d-1)|a|,
$$
the first equality is due to the fact that $\rho-\delta$ is traceless. We use the fact that $z$ is an eigenvalue of $\rho-\delta$ in the first inequality.

If $a<0$ and $d$ is odd, then $\prod_{i=1}^d(da-y_i)<0$, therefore $f((d-1)a)>0$. Meanwhile, $f(-\infty)<0$.

If $a<0$ and $d$ is even, then $\prod_{i=1}^d(da-y_i)>0$, therefore $f((d-1)a)<0$. Meanwhile, $f(-\infty)>0$.

In other words, $f((d-1)a)$ and $f(-\infty)$ have opposite signs when $a<0$. Hence, applying Bolzano's theorem to $f$, we have $f(t)=0$ for some $t\in(-\infty,(d-1)a)$, that is, there exists an eigenvalue $t<(d-1)a$ for $\rho-\delta$. Thus
$$
{\rm tr}|\rho-\delta|=-2(\lambda_{k+1}+\cdots+\lambda_d)\geq-2t>-2(d-1)a=2(d-1)|a|.
$$

On the other hand, if there exists $|x_i-\delta_i|\geq|da|$ for some $i$, according to lemma 1, the maximal singular value is larger than $d|a|$, hence the sum of the singular values of $\rho-\delta$ is larger than $2d|a|$, i.e., ${\rm tr}|\rho-\delta|\geq2d|a|>2(d-1)|a|$, where the first inequality is also due to the fact that $\rho-\delta$ is traceless.

In summary, in general ${\rm tr}|\rho-\delta|\geq2(d-1)|a|$. According to the above argument, the lower bound can be reached when $\delta={\rm diag}(\rho)$. Hence, $C_{\rm tr}(\rho)=2(d-1)|a|$, which prove our theorem.

Interestingly, the maximally coherence states $|\psi\ra=\displaystyle\frac{1}{\sqrt{d}}\sum_{i=1}^d|i\ra$ are included in the class of qudit states (\ref{rho}) \cite{baumgratz}. As an application, the maximally coherence states can be ordered by the coherence measures induced by trace distance, relative entropy and $l_1$ norm, that is,
$$
C_{\rm tr}(|\psi\ra)\leq C_{r}(|\psi\ra)\leq C_{l_1}(|\psi\ra)
$$
for $d\geq2$, where
$$
C_{\rm tr}(|\psi\ra)=\frac{2(d-1)}{d},\ \ C_{r}(|\psi\ra)=\log_2d,\ \ C_{l_1}(|\psi\ra)=d-1.
$$
It coincides with the fact that $C_{\rm tr}(|\phi\ra)\leq C_{l_1}(|\phi\ra)$ and $C_{r}(|\phi\ra)\leq C_{l_1}(|\phi\ra)$ for any pure states $|\phi\ra$ \cite{rana}.

According to the above theorem, now we can consider the monotonicity of the qudit states $\rho$ in (\ref{rho}) for average coherence under selective incoherent channels. However, only qubit states have the explicit expressions of the trace distance measure of coherence. And it is shown that the trace distance measure of coherence has the same expressions as the $l_1$ norm of coherence for qubits, so we will restrict the incoherent operators $K_n$ to $2\times d$ matrices. In addition, we suppose that not only the operators $K_n$ but also $K_n^\dagger$ is incoherent, that is, $K_n$ is strictly incoherent operator \cite{winter,yadin}. In fact, the strictly incoherent operators are included in the set of incoherent operators.

Let $K_n$ be a $2\times d$ matrix which is written as $K_n=([K_n]_{ij})$, where $i=1,\ 2,\ j=1,\ 2,\ \cdots,\ d$.
Let $\rho_n=\frac{K_n\rho K_n^\dagger}{p_n}$, where $p_n={\rm tr}(K_n\rho K_n^\dagger)$. Consider
\begin{eqnarray*}
&&\sum_n p_n C_{\rm tr}(\rho_n)=\sum_n p_n C_{l_1}(\rho_n)=\sum_n C_{l_1}(K_n\rho K_n^\dagger)\\
&=&|a|\sum_{k\neq l}\sum_n\sum_{i\neq j}|[K_n]_{ki}|\cdot|[K_n]_{lj}^*|\\
&\leq&|a|\sum_{k\neq l}\sum_n\sum_{i=1}^d|[K_n]_{ki}|\cdot\sum_{j=1}^d|[K_n]_{lj}^*|\\
&\leq&|a|\sum_{k\neq l}\sqrt{\sum_n\left(\sum_{i=1}^d|[K_n]_{ki}|\right)^2\cdot\sum_n\left(\sum_{j=1}^d|[K_n]_{lj}^*|\right)^2}
\end{eqnarray*}
here we use the Cauchy-Schwarz inequality in the final inequality.

In addition, due to the fact that $K_n^\dag{\cal I}K_n\subset{\cal I}$, i.e., $K_n$ is strictly incoherent operator, we get
\begin{eqnarray*}
\left(\sum_{i=1}^d|[K_n]_{ki}|\right)^2&=&\left(\sum_{i=1}^d|[K_n]_{ki}|\right)\cdot\left(\sum_{j=1}^d|[K_n]_{kj}^*|\right)\\
&=&\sum_{i,j=1}^d|[K_n]_{ki}\cdot[K_n]_{kj}^*|\\
&=&\sum_{i,j=1}^d|[K_n]_{ki}\cdot[K_n]_{kj}^*|\delta_{ij}\\
&=&\sum_{i=1}^d[K_n]_{ki}[K_n]_{ki}^*,
\end{eqnarray*}
such that
\begin{eqnarray*}
\sum_n p_n C_{\rm tr}(\rho_n)&\leq&|a|\sum_{k\neq l}\sqrt{\left(\sum_n\sum_{i=1}^d[K_n]_{ki}[K_n]_{ki}^*\right)\cdot\left(\sum_m\sum_{j=1}^d[K_m]_{lj}[K_m]_{lj}^*\right)}\\
&=&|a|\sqrt{\left(\sum_n\sum_{i=1}^d[K_n]_{1i}[K_n]_{1i}^*\right)\cdot\left(\sum_m\sum_{j=1}^d[K_m]_{2j}[K_m]_{2j}^*\right)}\\
&+&|a|\sqrt{\left(\sum_n\sum_{i=1}^d[K_n]_{2i}[K_n]_{2i}^*\right)\cdot\left(\sum_m\sum_{j=1}^d[K_m]_{1j}[K_m]_{1j}^*\right)}\\
&=&2|a|\sqrt{\left(\sum_n\sum_{i=1}^d[K_n]_{1i}[K_n]_{1i}^*\right)\cdot\left(\sum_m\sum_{j=1}^d[K_m]_{2j}[K_m]_{2j}^*\right)}\\
&\leq&|a|\left(\sum_n\sum_{i=1}^d[K_n]_{1i}[K_n]_{1i}^*+\sum_m\sum_{j=1}^d[K_m]_{2j}[K_m]_{2j}^*\right)\\
&=&|a|\sum_{i=1}^d\sum_n\sum_{k=1}^2[K_n]_{ki}^*[K_n]_{ki}=d|a|.
\end{eqnarray*}
It is noticed that the final equality is due to $\sum_n K_n^\dag K_n=I$. Following $d\leq2(d-1)$ when the dimension $d\geq2$, we obtain
$$
\sum_n p_n C_{\rm tr}(\rho_n)\leq d|a|\leq2(d-1)|a|=C_{\rm tr}(\rho),
$$
which prove the qudit states $\rho$ in (\ref{rho}) satisfy the condition (C2b) for the trace norm of coherence under strictly incoherent channels.

In conclusion, we derived an explicit expression for a class of qudit states under trace distance measure of coherence. Fortunately, it is found that the closest incoherent states for this class of qudits have the intuitively expected result, i.e., the closest incoherent state to the class of qudits is just the diagonal matrix obtained from it by deleting all off-diagonal elements. As an application, the maximally coherent states are ordered by the coherence measures induced by trace distance, relative entropy and $l_1$ norm. In addition, we conclude that the trace distance measure of coherence obeys the condition (C2b) for the class of qudit states if we restrict the strictly incoherent operators $K_n$ to $2\times d$ matrices. Whether the measure of coherence induced by the trace norm can be a good measure of coherence for general qudit states needs to further exploration.

\bigskip
\noindent{\bf Acknowledgments}\ We thank Mao-Sheng Li for his helpful discussions. The work is supported by NSF of China (Grant Nos. 11501247, 11504135),
NSF of Shandong Province (Grant No. ZR2014AL007) and a Project of Shandong Province Higher Educational Science and Technology
Program (Grant No. J14LI11). The work is also funded by Shandong Province College Young Teachers Visiting
Scholar Project.

\end{document}